\documentclass[a4paper,11pt]{article}
\usepackage{pos}

\title{Recent $J/\psi$ measurements at the PHENIX experiment}

\author*[a]{Xuan Li}
\author[]{On behalf of the PHENIX Collaboration}

\affiliation[a]{Los Alamos National Laboratory,\\
  P. O. Box 1663, Los Alamos, New Mexico, USA}

\emailAdd{xuanli@lanl.gov}

\abstract{Quarkonium is an ideal probe to explore the properties of quantum chromodynamics (QCD). Unlike Large Hadron Collider (LHC) measurements, quarkonium production at the Relativistic Heavy Ion Collider (RHIC) has different production mechanisms, can access different kinematic phase space and may experience different medium densities/temperatures. The PHENIX experiment has collected a large $J/\psi \to \mu^{+}\mu^{-}$ data set within its unique pseudorapidity region of $1.2<|\eta|<2.2$ in $p+p$, $p+A$ and $A+A$ collisions at $\sqrt{s}$ = 200 GeV from 2014 to 2016. Latest results of normalized charged particle multiplicity dependent normalized forward $J/\psi$ yields and normalized forward $\psi(2S)$ to $J/\psi$ ratio in 200 GeV $p+p$ collisions as well as forward $J/\psi$ azimuthal anisotropy in 200 GeV Au+Au collisions will be shown. Comparison with other RHIC and LHC measurements and latest theoretical calculations will be discussed. These PHENIX results provide their unique contributions in improving the understanding of the multi-parton interaction contribution to charmonium production and the recombination/coalescence contribution to charmonium formation within Quark Gluon Plasma (QGP) at RHIC energies.}

\FullConference{42nd International Conference on High Energy Physics (ICHEP2024)\\
18-24 July 2024\\
Prague, Czech Republic\\}

\begin{document}
\maketitle

\section{Introduction}
Although charmonium states are thought to be an ideal probe to test quantum chromodynamics (QCD) in high energy hadron-hadron and heavy-ion collisions, their production mechanism and how they interact with the surrounding nuclear medium is still lacking of sufficient knowledge. Creation of charmonia involves both perturbative and non-perturbative QCD aspects and initial- and/or final-state interaction effects. The PHENIX experiment at the Relativistic Heavy Ion Collider (RHIC) has good muon identification in the pseudorapidity region of $1.2<|\eta|<2.4$, which allows precise reconstruction of forward and backward $J/\psi$ via di-muon decay ($J/\psi \to \mu^{+}\mu^{-}$) in 200 GeV $p+p$, $p$+A and A+A collisions. The PHENIX experiment has collected 47~$pb^{-1}$ 200 GeV $p+p$ data in 2015 and 14.5~$nb^{-1}$ 200 GeV Au+Au data in 2014 and 2016. With these high statistics data sets at PHENIX, the first self-normalized event multiplicity dependent self-normalized forward $J/\psi$ yields as well as the forward $\psi(2S)$ to $J/\psi$ ratio have been measured in 200 GeV $p+p$ collisions. Moreover, the first forward $J/\psi$ azimuthal anisotropy measurement has been performed in 200 GeV Au+Au collisions. These PHENIX forward $J/\psi$ measurements provide further information in constraining the Multi-Parton Interactions (MPI) effects on the charmonium production and studying the creation and propagation of charmonia inside Quark Gluon Plasma (QGP).

\section{Event multiplicity dependent forward $J/\psi$ and $\psi(2S)$ measurements in 200 GeV $p+p$ collisions}
\begin{figure}[ht]
\centering
\includegraphics[width=0.49\textwidth,clip]{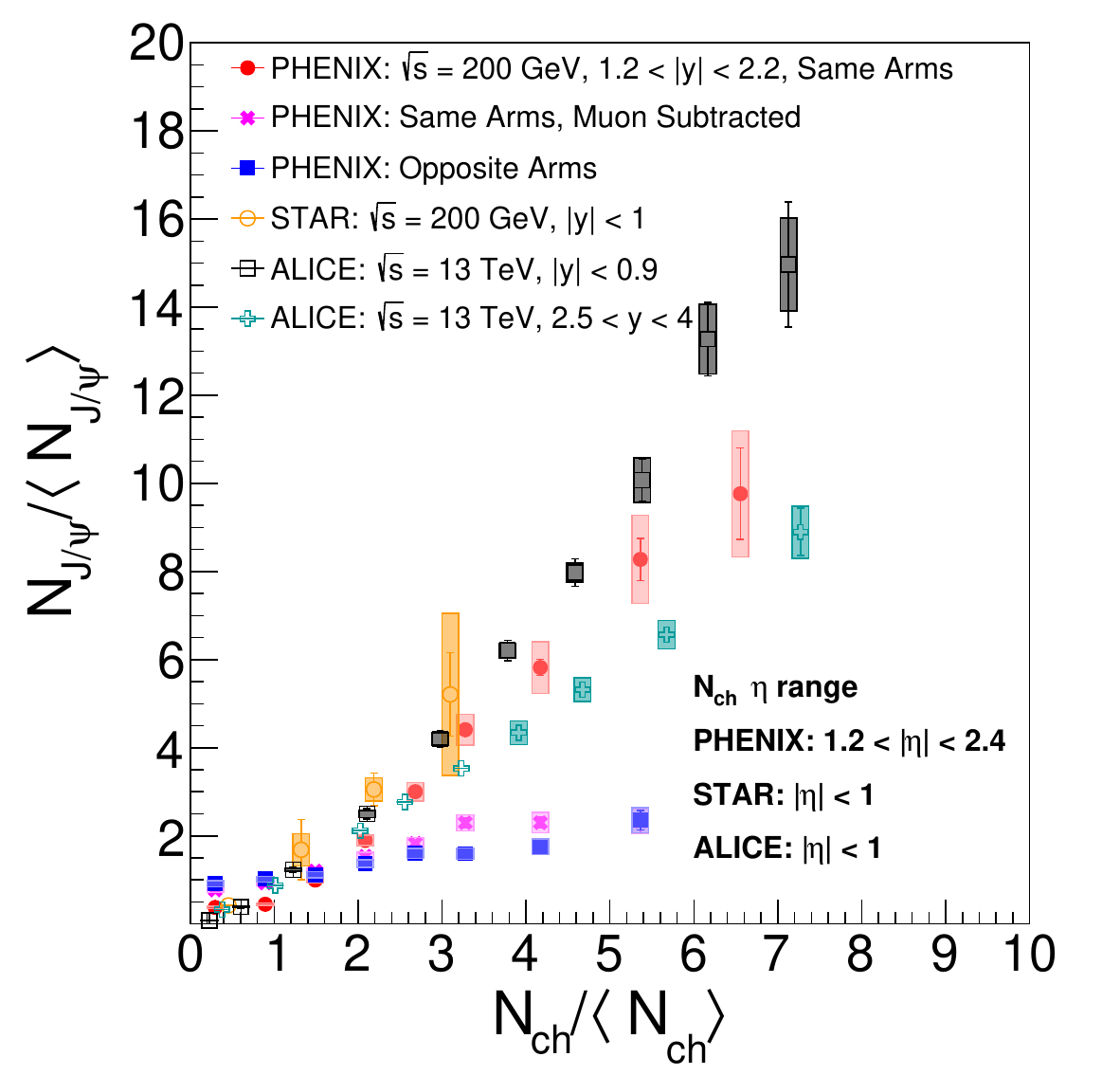}
\includegraphics[width=0.49\textwidth,clip]{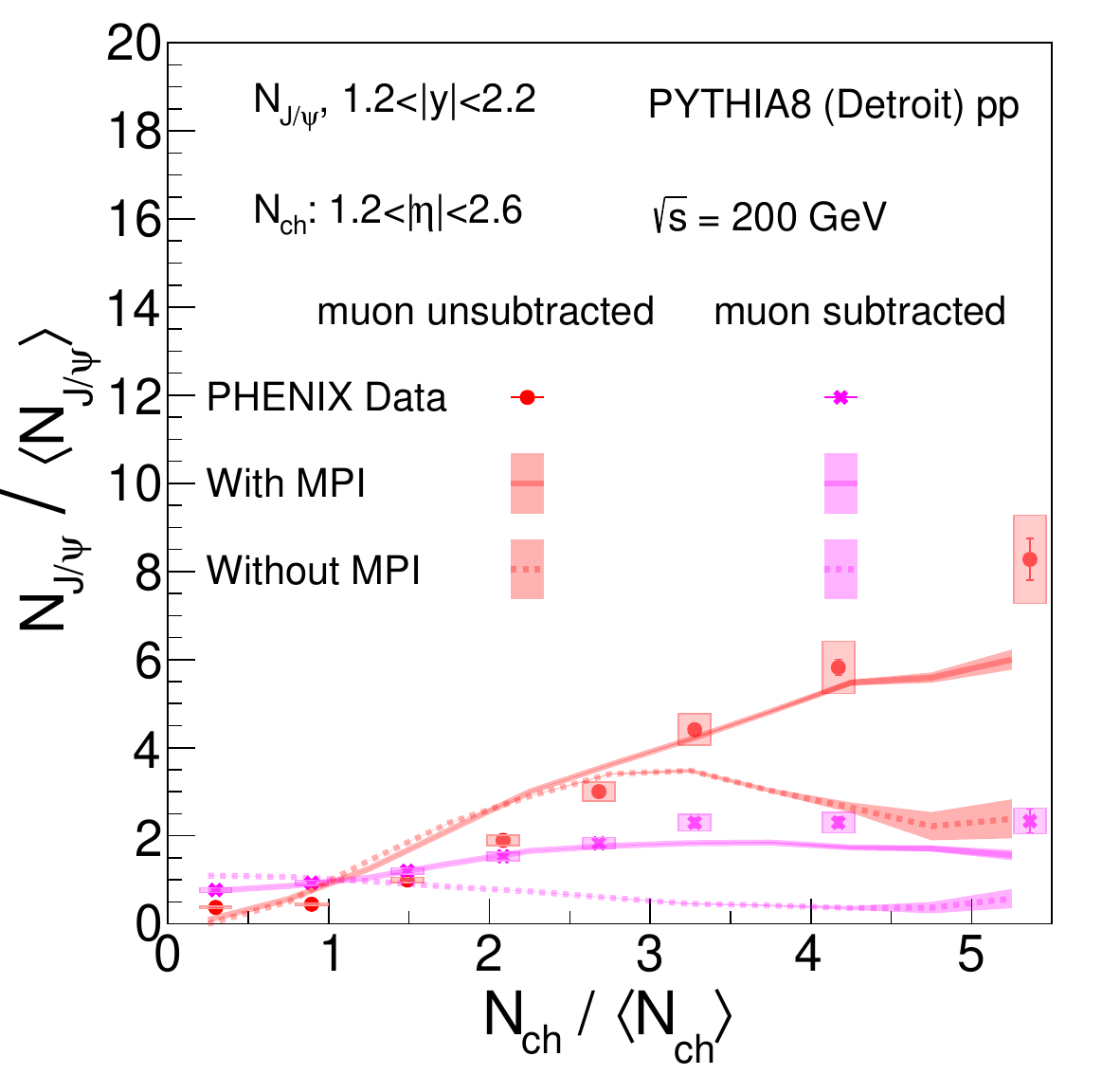}
\caption{Left: the self-normalized event multiplicity ($N_{ch}/\langle N_{ch}\rangle$) dependent self-normalized $J/\psi$ yields ($N_{J/\psi}/\langle N_{J/\psi}\rangle$) in $p+p$ collisions at $\sqrt{s}=200$~GeV at PHENIX in comparison with STAR and ALICE results. The PHENIX results with $J/\psi$ and charged particles measured in the same (opposite) pseudorapidity region are shown in red (blue). The PHENIX measurement with the event multiplicity after subtracting the $J/\psi$ decayed muons are shown in magenta. Right: comparison of PHENIX results and PYTHIA8 (Detroit) calculations for the $N_{ch}/\langle N_{ch}\rangle$ dependent $N_{J/\psi}/\langle N_{J/\psi}\rangle$ with and without the subtraction of $J/\psi$ decay muons in the event multiplicity. The distributions without (with) the muon subtraction are shown in red (magenta). PYTHIA8 (Detroit) calculations are divided into with and without MPI effects.}
\label{fig-1}       
\end{figure}

Recent measurements of self-normalized $J/\psi$ yields as a function of self-normalized event multiplicity at RHIC and the LHC in $p+p$ collisions \cite{star_pp, alice_pp1,alice_pp2} present an increasing trend of $J/\psi$ yields with the event multiplicity. These results suggest that the MPI effects should not be neglected in the $J/\psi$ production. Good $J/\psi$ and $\psi(2S)$ signals have been measured in the pseudorapidity region of $1.2<|\eta|<2.4$ at PHENIX in 200 GeV $p+p$ collisions with 47~$pb^{-1}$ integrated luminosity of the 2015 data set. The analyzed events were selected by the Minimum Bias (MB) trigger and a dimuon trigger, which requires at least two muon tracks were measured in the forward and backward PHENIX MuID detector. The PHENIX experiment has measured the self-normalized event multiplicity ($N_{ch}/\langle N_{ch}\rangle$) dependent self-normalized forward $J/\psi$ yields ($N_{J/\psi}/\langle N_{J/\psi}\rangle$) in three different scenarios \cite{jpsi_pp} as shown in the left panel of Figure~\ref{fig-1}. In the self-normalized event multiplicity observable: $N_{ch}/\langle N_{ch}\rangle$, $N_{ch}$ represents the number of charged particles detected by the PHENIX FVTX detector in the pseudorapidity regions of $1.2<\eta<2.4$ (forward) and $-2.4<\eta<-1.2$ (backward), and $\langle N_{ch}\rangle$ is referred to the average $N_{ch}$ in MB events. The self-normalized $J/\psi$ yields, denoted as $N_{J/\psi}/\langle N_{J/\psi}\rangle$, are measured by the PHENIX muon arms covering $1.2<y<2.2$ (forward) and $-2.2<y<-1.2$ (backward) respectively. Similar multiplicity dependence as the STAR mid-rapidity $J/\psi$ measurement \cite{star_pp} has been observed by the PHENIX forward $J/\psi$ results when $J/\psi$ and charged particles are measured in the same pseudorapidity region. Reduced dependence on the self-normalized event multiplicity can be achieved by either measuring the $J/\psi$ and charged particles with a large pseudorapidity gap (e.g., in opposite pseudorapidity regions) or through subtracting $J/\psi$ decay muons in the event multiplicity calculations.

\begin{figure}[ht]
\centering
\includegraphics[width=0.99\textwidth,clip]{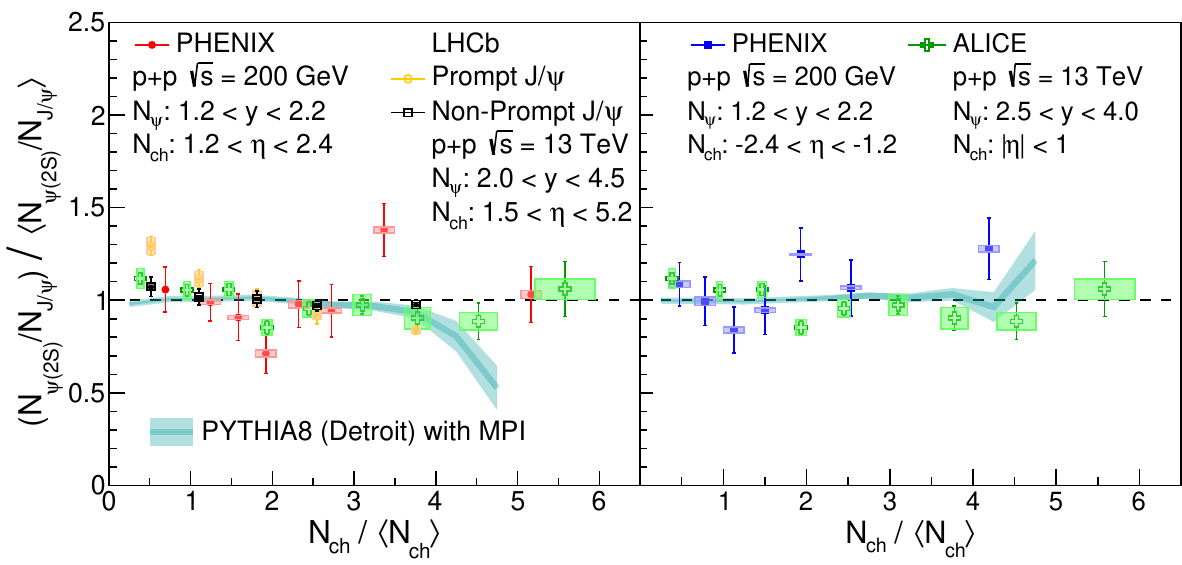}
\caption{The self-normalized event multiplicity ($N_{ch}/\langle N_{ch}\rangle$) dependent self-normalized $\psi(2S)$ to $J/\psi$ ratio ($(N_{\psi(2S)}/N_{J/\psi})/\langle N_{\psi(2S)}/N_{J/\psi} \rangle$) in $p+p$ collisions at $\sqrt{s}=200$~GeV at PHENIX in comparison with both the PYTHIA8 (Detroit) calculations with MPI effects and the LHCb and ALICE results. Left: both the $J/\psi$ and charged particles are measured in the same pseudorapidity region at PHENIX (red closed circles) and LHCb (yellow/black open markers). Right: there are rapidity gaps between $J/\psi$ and charged particles in the $N_{ch}/\langle N_{ch}\rangle$ dependent $(N_{\psi(2S)}/N_{J/\psi})/\langle N_{\psi(2S)}/N_{J/\psi} \rangle$ measurements at PHENIX (blue closed rectangulars) and ALICE (green open crossings).}
\label{fig-2}       
\end{figure}

To have a better understanding about the effects on the event multiplicity dependent $J/\psi$ yields, comparison between the PHENIX measurements and PYTHIA8 Detroit calculations with the same kinematic selections has been performed. As shown in the right panel of Figure~\ref{fig-1},  the PYTHIA8 Detroit tuned with the MPI effects \cite{py8_detroit} agrees with the PHENIX measurements within $\sim 1\sigma$ uncertainties. However the PYTHIA8 Detroit calculations without including the MPI effects fail to reproduce the distributions measured in data. Therefore the MPI effects should be included to accurately model the $J/\psi$ production in 200 GeV $p+p$ collisions. There is smaller dependence on the event multiplicity for forward $J/\psi$ production at PHENIX than that at ALICE, which indicates that the MPI effects would be stronger in $p+p$ collisions at the LHC energies.

The PHENIX experiment also measured the self-normalized event multiplicity dependent self-normalized $\psi(2S)$ to $J/\psi$ ratio ($(N_{\psi(2S)}/N_{J/\psi})/\langle N_{\psi(2S)}/N_{J/\psi} \rangle$) at forward rapidity in 200 GeV $p+p$ collisions to investigate potential final-state interaction effects on the charmonium production. The left panel of Figure~\ref{fig-2} presents the PHENIX $N_{ch}/\langle N_{ch} \rangle$ dependent $(N_{\psi(2S)}/N_{J/\psi})/\langle N_{\psi(2S)}/N_{J/\psi} \rangle$ at forward rapidity in 200 GeV $p+p$ collisions in comparison with PYTHIA8 Detroit tuned with MPI effects and LHCb results of prompt and non-prompt $\psi(2S)$ to $J/\psi$ ratio measured at forward rapidity \cite{lhb_pp}. The right panel of Figure~\ref{fig-2}  shows the PHENIX measurements of self-normalized event multiplicity dependent forward $\psi(2S)$ to $J/\psi$ ratio in comparison with PYTHIA8 Detroit calculations including the MPI effects and ALICE results \cite{alice_pp3} when $J/\psi$ ($\psi(2S)$) and charged particles are measured with a rapidity gap. These $N_{ch}/\langle N_{ch} \rangle$ dependent $(N_{\psi(2S)}/N_{J/\psi})/\langle N_{\psi(2S)}/N_{J/\psi} \rangle$ measurements are consistent with unity within $\sim 1\sigma$ uncertainties. No significant dependence on the event multiplicity and different rapidity gaps between $J/\psi$ ($\psi(2S)$) and charged particles is found. Calculations of PYTHIA8 Detroit tuned with MPI effects present similar results of the self-normalized event multiplicity dependent $(N_{\psi(2S)}/N_{J/\psi})/\langle N_{\psi(2S)}/N_{J/\psi} \rangle$ as the PHENIX measurements. These studies prefer that there are no significant final-state interaction effects on the charmonium production in $p+p$ collisions at $\sqrt{s}=200$~GeV.

\section{Forward $J/\psi$ azimuthal anisotropy measurements in 200 GeV Au+Au collisions}
The PHENIX experiment has measured the first azimuthal anisotropy ($v_{2}$) of inclusive $J/\psi$ at forward rapidity ($1.2<|y|<2.2$) in 200 GeV Au+Au collisions from the 2014 and 2016 run \cite{jpsi_v2}. The $J/\psi$ $v_{2}$ is determined by the event plane method. To maximize potential $v_{2}$ signal and mitigate event selection biases, these measurements are carried out with several selections of collision centrality: $0\%-50\%$, $10\%-60\%$, and $10\%-40\%$. Clear $J/\psi$ signals have been found in these near-central Au+Au collisions with the reconstructed $J/\psi$ $p_{T}$ measured down to zero. The total integrated luminosity is 14.5 $nb^{-1}$, which represents the largest data set of Au+Au collisions collected at PHENIX.

\begin{figure}[ht]
\centering
\includegraphics[width=0.99\textwidth,clip]{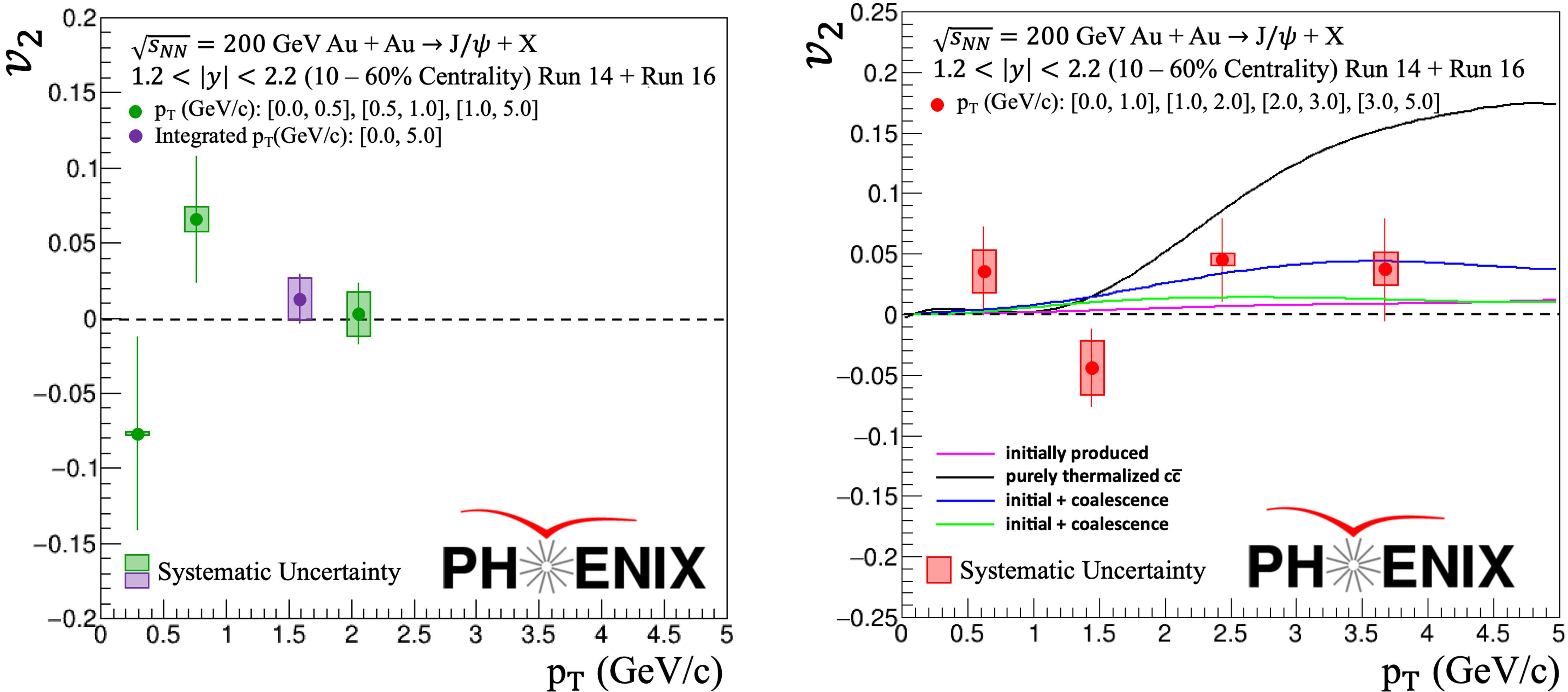}
\caption{$J/\psi$ azimuthal anisotropy ($v_{2}$) measured at PHENIX with $1.2<|y|<2.2$ and $0<p_{T}<5$~ GeV/c in 10-60$\%$ 200 GeV Au+Au collisions. Left: comparison between $p_{T}$ dependent and $p_{T}$ integrated forward $J/\psi$ $v_{2}$ measured at PHENIX. Right: comparison between $p_{T}$ dependent forward $J/\psi$ $v_{2}$ with various theoretical model predictions \cite{th1,th2,th3,th4}.}
\label{fig-3} 
\end{figure}

The left panel of Figure~\ref{fig-3} presents the PHENIX $p_{T}$ integrated (purple) and $p_{T}$ dependent (green) forward $J/\psi$ $v_{2}$ within the region of $0<p_{T}<5$~ GeV/c and $1.2<|y|<2.2$ in $10\%-60\%$ 200 GeV Au+Au collisions. Both $J/\psi$ $v_{2}$ results measured at PHENIX are consistent with zero within uncertainties, which are similar to the STAR result of midrapidity $J/\psi$ $v{2}$ for the centrality of $10\%-60\%$ in Au+Au collisions at $\sqrt{s_{NN}} =$ 200 GeV \cite{star_jpsi_v2}. However, these RHIC measurements are systematically lower than the ALICE forward $J/\psi$ $v{2}$ measured in $10\%-50\%$ in Pb+Pb collisions at $\sqrt{s_{NN}} =$ 5 TeV \cite{alice_jpsi_v2}. To properly describe the $J/\psi$ azimuthal anisotropy in heavy ion collisions, the interplay between the "initial produced $c\bar{c}$" and "coalescence from thermalized $c\bar{c}$" should be included in theoretical models. The right panel of Figure~\ref{fig-3} compares the $p_{T}$ dependent forward $J/\psi$ $v{2}$ measured at PHENIX with different model calculations \cite{th1,th2,th3,th4}, which were originally used to interpret the STAR midrapidity $J/\psi$ $v_{2}$ measurements. The PHENIX result can be well described by theory calculations which include coalescence of partially thermalized $c\bar{c}$. However, it is challenging to distinguish whether the non-significant forward $J/\psi$ $v_{2}$ measured at PHENIX is driven by only primordial $J/\psi$ production through initial hard scattering or hadronization dominated by coalescence. The clear difference between forward $J/\psi$ $v_{2}$ at RHIC and that measured at the LHC indicates that the $J/\psi$ production mechanism in heavy ion collisions could vary with the collision energy.

\section{Summary}
The PHENIX experiment has just released several first forward $J/\psi$ and $\psi(2S)$ measurements at RHIC. The self-normalized event multiplicity dependent $J/\psi$ yields at forward rapidity in 200 GeV $p+p$ collisions are consistent with the STAR and ALICE results when $J/\psi$ and charged particles are measured in the same pseudorapidity region. Results of the event multiplicity dependent self-normalized $J/\psi$ before and after subtracting $J/\psi$ decay muons in the event multiplicity counting can be well described by the PYTHIA8 Detroit tuned with the MPI effects. The self-normalized $\psi(2S)$ to $J/\psi$ ratio measured at forward rapidity in 200 GeV $p+p$ collisions is in agreement with unity within uncertainties and is also consistent with other results with different rapidity gaps and at different collisions energies. These results reflect that the MPI effects should be included in models to properly describe the $J/\psi$ production in $p+p$ collisions at $\sqrt{s}=200$~GeV and final-state interaction effects do not play a significant role in the charmonium production in $p+p$ collisions. The PHENIX measurement of forward $J/\psi$ azimuthal anisotropy in near-central 200 GeV Au+Au collisions is consistent with zero and the midrapidity $J/\psi$ $v_{2}$ measured at STAR. This suggests that the $J/\psi$ $v_{2}$ measured at RHIC could experience little rapidity dependence.

\end{document}